\documentclass[twocolumn,showpacs,preprintnumbers,prd]{revtex4-1}
\usepackage{psfrag,graphicx,epsfig,amsmath,amssymb,bm,tikz}
\usetikzlibrary{arrows,decorations.markings,decorations.pathmorphing}

\newcommand{\ord}{{\cal O}}
\newcommand{\Mkk}{M_{\rm KK}}

\makeatletter
\DeclareRobustCommand*{\bfseries}{%
  \not@math@alphabet\bfseries\mathbf
  \fontseries\bfdefault\selectfont
  \boldmath
}
\makeatother

\begin{document}

\preprint{MZ-TH/11-30} 

\title{A Solution to the Flavor Problem of Warped Extra-Dimension Models}

\author{Martin Bauer, Raoul Malm and Matthias Neubert} 

\affiliation{Institut f\"ur Physik, Johannes Gutenberg-Universit\"at,
D-55099 Mainz, Germany}

\date{October 3, 2011}

\begin{abstract}
\noindent
A minimal solution to the flavor problem of warped extra-dimension models, i.e.\ the excessive mixed-chirality contribution to CP violation in $K$--$\bar K$ mixing arising from Kaluza-Klein (KK) gluon exchange, is proposed. Extending the strong-interaction gauge group in the bulk by an additional $SU(3)$, and breaking this symmetry to QCD via boundary conditions, the constraints arising from the $\epsilon_K$ parameter are significantly relaxed. As a result, KK scales $\Mkk\sim 2$\,TeV are consistent with all flavor observables without significant fine-tuning. The model predicts the existence of an extended Higgs sector featuring massive color-octet scalars and a tower of KK pseudo-axial gluon resonances, whose existence is not in conflict with recent LHC dijet bounds.
\end{abstract}


\maketitle

Randall-Sundrum (RS) models featuring a warped extra dimension, in which Minkowski space is embedded in a slice of five-dimensional (5D) anti-de Sitter space with curvature $k$ and radius $r$, represent one of the most compelling candidates for an extension of the Standard Model (SM) \cite{Randall:1999ee}. These models can explain the hierarchy problem and the flavor puzzle in terms of the same geometrical mechanism. At different points along the extra dimension, the warp factor $e^{-kr|\phi|}$ rescales the units of length and mass in such a way that the fundamental Planck scale is warped down from $10^{19}$\,GeV to the TeV scale as $\phi$ varies between 0 (UV brane) to $\pi$ (IR brane). As a consequence, the Higgs mass is stabilized around the electroweak scale as long as the Higgs sector is localized near the IR brane. All other SM particles are allowed to propagate in the bulk. Depending on the choice of the 5D mass parameters, the wave functions of the lowest-lying (SM) fermions turn out to be localized near one of the two branes \cite{Grossman:1999ra,Gherghetta:2000qt}. Their overlap with the Higgs profile determines the masses of the SM fermions and their Yukawa couplings. Fermion localization in the warped extra dimension naturally explains the hierarchies observed in the spectrum of quark masses and mixing angles \cite{Huber:2003tu}. The same flavor-dependent overlap factors govern the couplings of the SM quarks to Kaluza-Klein (KK) excitations of the gauge bosons, which like the Higgs boson are localized near the IR brane. As a result, tree-level flavor-changing neutral current (FCNC) processes induced by KK boson or Higgs exchange are strongly suppressed. This so-called RS-GIM mechanism \cite{Agashe:2004ay} has proven to be very effective in suppressing FCNC transitions in $B$, $D$ and kaon physics \cite{Huber:2003tu,Agashe:2004ay,Moreau:2006np,Csaki:2008zd,Casagrande:2008hr,Bauer:2009cf}. The single exception is the CP-violating observable $\epsilon_K$ in $K$--$\bar K$ oscillations \cite{Csaki:2008zd}, for which a fine-tuning of order 1\% is required for KK masses in the few TeV range. The reason lies in the strong chiral enhancement of mixed-chirality four-quark operators contributing to this observable. 

Different proposals to mitigate this ``RS flavor problem'' have been made. For example, imposing a discrete flavor symmetry in the bulk one can eliminate the relevant flavor non-diagonal overlap integrals at leading order in $v^2/\Mkk^2$ \cite{Cacciapaglia:2007fw,Fitzpatrick:2007sa,Santiago:2008vq,Csaki:2008eh}. The problem that this mechanism is very sensitive to small symmetry-breaking effects can be avoided by gauging the flavor symmetry. While the appropriate implementation of such a symmetry gives rise to viable models \cite{Csaki:2009wc}, the attractive feature of a purely geometric explanation of flavor hierarchies is lost if a flavor symmetry is imposed in the 5D theory. An alternative is to lower the UV cutoff of the model to a value significantly below the Planck scale \cite{Davoudiasl:2008hx}. This rescales the Wilson coefficients of the dangerous operators proportional to $L=\ln(\Lambda_{\rm UV}/\Lambda_{\rm IR})$. However, it turns out that for too small $L$ values the RS GIM mechanism becomes inoperative \cite{Bauer:2008xb}, and therefore the RS flavor problem reappears. The $\epsilon_K$ constraint can also be relaxed by taking the Higgs sector off the IR brane and including loop corrections in the relation between the 5D and 4D gauge couplings \cite{Agashe:2008uz}.

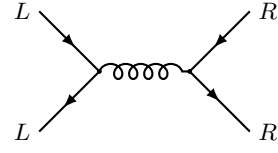
\begin{figure}
\begin{center}
\begin{tikzpicture}[scale=0.8]
\draw[thick,-latex] (-0.5,1) to (0.1,0.4);
\draw[thick] (0,0.5) to (0.5,0);
\draw[thick] (-0.5,-1) to (0,-0.5);
\draw[thick,latex-] (-0.1,-0.6) to (0.5,0);
\draw[decoration={coil,aspect=1,segment length=2mm,amplitude=1mm},decorate,thick] (0.5,0) to (2,0);
\draw[thick] (2,0) to (2.5,.5);
\draw[thick,latex-] (2.4,.4) to (3,1);
\draw[thick,-latex] (2,0) to (2.5,-.5);
\draw[thick] (2.4,-.4) to (3,-1);
\fill[color=black!] (0.5,0) circle (0.4mm);
\fill[color=black!] (2,0) circle (0.4mm);
\node[left] at (-.5,1){$L$};
\node[left] at (-.5,-1){$L$};
\node[right] at (3,-1){$R$};
\node[right] at (3,1){$R$};
\end{tikzpicture}
\end{center}
\vspace{-3mm}
\caption{Tree-level contributions to mixed-chirality four-quark operators due to KK gluon and pseudo-axial gluon exchange.}
\label{fig:graph}
\end{figure}

In this Letter, we propose a novel solution to the RS flavor problem based on an extension of the strong-interaction gauge group in the bulk. The enlarged symmetry is broken via boundary conditions (BCs) to the gauge group of the SM. We show that the dangerous mixed-chirality operators contributing to $\epsilon_K$ are cancelled for a large class of BCs. More specifically, we assume that the QCD gauge symmetry in the bulk is extended to $SU(3)_D\times SU(3)_S$. The 5D color-octet gauge bosons $G_\mu^D$ and $G_\mu^S$ couple to $SU(2)_L$ quark doublets and singlets with coupling strengths $g_D$ and $g_S$, respectively:
\begin{equation}\label{sc1:1}
   {\cal L}_{\rm int}
   \ni g_D\,\bar Q\,G^D_\mu\gamma^\mu Q + g_S\,\bar q\,G^S_\mu\gamma^\mu q \,.
\end{equation}
We break this extended symmetry by a suitable choice of BCs on the UV and IR branes. In order to recover the SM gluon, the couplings of the combination $g_\mu=G_\mu^D\cos\theta+G_\mu^S\sin\theta$, with $\tan\theta=g_D/g_S$, to 5D doublets and singlets must be identified with the strong coupling $g_s$, and the corresponding KK tower must allow for a zero mode. We will refer to the orthogonal combination $A_\mu=-G_\mu^D\sin\theta+G_\mu^S\cos\theta$ as \emph{pseudo-axial gluon}. Eq.~\eqref{sc1:1} implies
\begin{equation}\label{sc1:2}
\begin{aligned}
   {\cal L}_{\rm int}
   &\ni g_s \left(\bar Q\,g_\mu\gamma^\mu Q + \bar q\,g_\mu\gamma^\mu q \right) \\ 
   &\mbox{}+ g_s \left( -\tan\theta\,\bar Q\,A_\mu\gamma^\mu Q 
    + \cot\theta\,\bar q\,A_\mu\gamma^\mu q \right) ,
\end{aligned}
\end{equation}
where $g_s=\sqrt{g_D^2+g_S^2}\,\sin\theta\cos\theta$. Up to higher-order corrections in $v^2/\Mkk^2$, the left-handed (right-handed) SM quarks are the zero modes of the 5D doublets $Q$ (singlets $q$). As a result, the contributions to mixed-chirality four-quark operators induced by KK gluon and pseudo-axial gluon exchange (see Figure~\ref{fig:graph}) cancel each other, independently of the mixing angle $\theta$, provided that the corresponding 5D propagators have the same form.

In Feynman gauge, the propagator for a 5D gauge field in the mixed position-momentum representation can be written as
\begin{equation}
   i D(t,t';p)\,g^{\mu\nu}
   = \sum_n\frac{-i g^{\mu\nu}}{p^2-m_n^2+i\epsilon}\,\chi_n(t)\,\chi_n(t') \,,
\end{equation}
where the right-hand side consists of the sum over the 4D propagators for the KK modes, and the profiles $\chi_n(t)$ describe the KK wave functions along the extra dimension. We parametrize the fifth coordinate by a dimensionless variable $t=\epsilon\,e^{kr|\phi|}\in[\epsilon, 1]$, where $\epsilon\sim M_W/M_{\rm Pl}$ is the ratio of the electroweak and Planck scales. The two branes are localized at $t=\epsilon$ (UV) and $t=1$ (IR). In low-energy processes the propagator is required at $p^2\approx 0$, in which case the sum over KK modes can be evaluated in closed form. The result depends on the BCs imposed on the wave functions $\chi_n(t)$. For the KK tower of the gluon, the choice of Neumann-Neumann BCs, $\partial_t\chi_n^{(g)}(\epsilon)=\partial_t\chi_n^{(g)}(1)=0$, allows for a massless mode. In this case the sum over KK modes gives (ignoring tiny $\ord(\epsilon^2)$ terms) \cite{Casagrande:2008hr}
\begin{equation}\label{KKsum}
\begin{aligned}
   &\sum_{n\ge 1}\,\frac{\chi_n^{(g)}(t)\,\chi_n^{(g)}(t')}{m_n^2}
   = \frac{L}{4\pi\Mkk^2} \\
   &\times \left[ t_<^2 - \frac{t^2}{L} \left( \frac12 - \ln t \right)
    - \frac{t'^2}{L} \left( \frac12 - \ln t' \right) + \frac{1}{2L^2} \right] ,
\end{aligned}
\end{equation}
where $t_<^2\equiv\mathrm{min}(t^2,t'^2)$, $L=\ln(1/\epsilon)$, and $\Mkk=k\epsilon$ sets the mass scale for the low-lying KK modes. Only the first term on the right-hand side ($\propto t_<^2$) gives rise to $\Delta F=2$ transitions such as meson-antimeson mixing. A general parametrization of new-physics contributions to the effective Hamiltonian for $K$--$\bar K$ mixing in the RS model requires the operators 
\begin{equation}
\begin{aligned}
   Q_1^{sd} &= (\bar d_L\gamma^\mu s_L )\,(\bar d_L\gamma_\mu s_L ) \,, \\
   \widetilde Q_1^{sd} &= (\bar d_R\gamma^\mu s_R )\,(\bar d_R\gamma_\mu s_R) \,, \\
   Q_4^{sd} &= -\frac12\,(\bar d_L^\alpha\gamma^\mu s_L^\beta)\,
    (\bar d_R^\beta\gamma_\mu s_R^\alpha) 
    = (\bar d_R s_L)\,(\bar d_L s_R) \,, \\   
   Q_5^{sd} &= -\frac12\,(\bar d_L\gamma^\mu s_L)\,(\bar d_R\gamma_\mu s_R) 
    = (\bar d_R^\alpha s_L^\beta)\,(\bar d_L^\beta s_R^\alpha) \,,
\end{aligned}
\end{equation}
multiplied by Wilson coefficients $C_{1,4,5}$ and $\widetilde C_1$. In the SM only $Q_1^{sd}$ contributes, while in the minimal RS model all four Wilson coefficients receive contributions of order $g_s^2 L/\Mkk^2$ times small flavor-changing couplings from KK gluon exchange \cite{Bauer:2009cf}. The suppression by the KK scale $\Mkk\sim\mbox{few TeV}$ is compensated by the fact that in the RS model $K$--$\bar K$ mixing arises at tree level, while it is loop-suppressed in the SM. In terms of the effective Hamiltonian, the $K_L-K_S$ mass difference $\Delta m_K$ and the CP-violating quantity $\epsilon_K$ are given by
\begin{equation}
\begin{aligned}
   \Delta m_K &= 2\,\mbox{Re}\,\langle K^0|{\cal H}_{\rm eff}^{\Delta S=2}|\bar K^0\rangle \,, \\
   \epsilon_K &= \frac{\kappa_\epsilon\,e^{i\phi_\epsilon}}%
                      {\sqrt{2}\left(\Delta m_K\right)_{\mathrm{exp}}}\,
    \mbox{Im}\,\langle K^0|{\cal H}_{\rm eff}^{\Delta S=2}|\bar K^0\rangle \,,
\end{aligned}
\end{equation}
where $\phi_\epsilon=(43.51\pm 0.05)^\circ$ and $\kappa_\epsilon=0.92\pm 0.02$ \cite{Buras:2008nn}. In terms of the Wilson coefficients renormalized at the KK scale, one obtains approximately \cite{Bauer:2009cf}
\begin{equation}\label{roughly}
   \langle K^0|{\cal H}_{\rm eff}^{\Delta S=2}|\bar K^0\rangle
   \propto C_1 + \widetilde C_1 + 115 \left( C_4 + \frac{C_5}{N_c} \right) .
\end{equation}
The key role of the mixed-chirality operators $Q_{4,5}^{sd}$ results from a chiral enhancement $\sim(m_K^2/m_s)^2$ and a significant renormalization-group evolution from $\Mkk$ to a hadronic scale $\mu\approx 2$\,GeV. These contributions tend to dominate over the SM unless the KK scale is raised above 10\,TeV. 

In order to cancel the contributions from KK gluon exchange to the mixed-chirality operators, the sum over the pseudo-axial gluon tower must give rise to the same $\Delta F=2$ contributions as in the case of (\ref{KKsum}), up to corrections of order $v^2/\Mkk^4$. Remarkably, we find that this is indeed the case for a very large class of BCs. For example, imposing Neumann BCs $\partial_t\chi_n^{(A)}(\epsilon)=0$ on the UV brane and mixed BCs
\begin{equation}\label{a1}
   \partial_t\chi_n^{(A)}(1) = - r_1\,\chi_n^{(A)}(1)
\end{equation}
on the IR brane, we obtain
\begin{equation}\label{KKsum12}
   \sum_n \frac{\chi_n^{(A)}(t)\,\chi_n^{(A)}(t')}{m_n^2} 
   = \frac{L}{4\pi\Mkk^2} \left[ t_<^2 - t^2 - t'^2 + 1 + \frac{2}{r_1} \right] \!.
\end{equation}
Dirichlet BCs, $\chi_n^{(A)}(1)=0$, correspond to the limit $r_1\to\infty$. For $t=t'$ each term in the sum is positive, and the requirement of positivity of the right-hand side (for $\epsilon\le t\le 1$) implies that $r_1\ge 0$; for negative $r_1$, the BCs (\ref{a1}) would lead to tachyonic solutions with $m_n^2<0$ for some $n$. The leading $\Delta F=2$ contributions arise from the $t_<^2$ term, which has the same coefficient as in (\ref{KKsum}). Dropping $\ord{(\epsilon^2)}$ corrections, the mass $m_{A_1}$ of the lightest pseudo-axial gluon is determined by the smallest eigenvalue of the equation $x_1\,[Y_0(\epsilon x_1)\,J_0(x_1)-Y_0(x_1)]=-r_1\,[Y_0(\epsilon x_1)\,J_1(x_1)-Y_1(x_1)]$, with $x_1=m_{A_1}/\Mkk$. For values $r_1\ll 1$ we obtain $x_1\approx\sqrt{r_1/L}$, whereas for $r_1\gg 1$ the solution quickly approaches $x_1\approx 0.235$ (for $\epsilon=10^{-16}$). Unfortunately, such a light pseudo-axial gluon, which would couple with almost equal strength to all SM quarks, is excluded by LHC dijet bounds \cite{Khachatryan:2010jd,Aad:2011aj}, unless the KK scale is raised above 10\,TeV, in which case the RS flavor problem does not pose itself in the first place.

We are thus led to consider mixed boundary conditions 
\begin{equation}\label{aeps}
   \partial_t\chi_n^{(A)}(\epsilon) = r_\epsilon\,\chi_n^{(A)}(\epsilon) 
\end{equation}
also on the UV brane, where once again we must impose $r_\epsilon>0$. This gives 
\begin{equation}\label{KKsum12b}
   \sum_n\frac{\chi_n^{(A)}(t)\,\chi_n^{(A)}(t')}{m_n^2} 
   = \frac{L}{4\pi\Mkk^2}\!\left[ t_<^2 - \frac{r_1}{2+r_1}\,t^2 t'^2 + \ord{(\epsilon)} \right]
\end{equation}
independently of the value of $r_\epsilon$, as long as we assume that $r_\epsilon$ is not of order $\epsilon$, which would appear as a rather artificial choice. While the coefficient of the $t_<^2$ term is still the same as in (\ref{KKsum}), the new contribution proportional to $t^2 t'^2$ gives rise to additional $\Delta F=2$ effects. In order to make these contributions subleading, we need to require that $r_1=\ord{(v^2/\Mkk^2)}$ or smaller. In this case, the mass of the lightest pseudo-axial gluon is determined by the first eigenvalue of the equation $x_1 J_0(x_1)=-r_1 J_1(x_1)$, and hence $m_{A_1}\approx(2.405+r_1/2.405)\,\Mkk$ to first order in $r_1$. Since the wave functions of the KK pseudo-axial gluons are localized near the IR brane, they couple primarily to top-quarks. Even for KK scales as low as 1\,TeV, such resonances are not excluded by existing collider data. 

The mixed BC (\ref{a1}) with small $r_1$ is indeed natural. In order to write down gauge-invariant Yukawa couplings for the quarks in our extended model, we must extend the Higgs sector as well. A minimal possibility is that the scalar field coupling to quarks transforms as $(\bm{3},\overline{\bm{3}},\bm{2})$ under $SU(3)_D\times SU(3)_S\times SU(2)_L$ \cite{Frampton:1987dn}. In the process of electroweak symmetry breaking, the color-singlet vacuum expectation value of the Higgs field  automatically breaks $SU(3)_D\times SU(3)_S\to SU(3)_c$ on the IR brane. While the precise nature of this symmetry breaking is model dependent, the connection with the weak scale naturally implies that $r_1=\xi L\,v^2/\Mkk^2$ with $\xi={\cal O}(1)$, where a factor of $L$ appears generically due to our choice of BCs.

The $\Delta F=2$ contributions to the mixed-chirality operators cancel at leading order in $1/\Mkk^2$ when one adds the effects from the KK towers of gluons and pseudo-axial gluons, independently of the value of $\theta$. The coefficients $C_1$ and $\widetilde C_1$ become enhanced by $\theta$-dependent factors with respect to the minimal RS model, while the coefficients $C_{4,5}\sim v^2/\Mkk^4$ become sufficiently suppressed to offset the large prefactor in (\ref{roughly}). Using the notation from \cite{Casagrande:2008hr,Bauer:2009cf} for the relevant overlap integrals, we obtain for the leading terms (at a scale $\mu\approx\Mkk$) 
\begin{eqnarray}
   C_1 &=& \frac{\pi\alpha_s L}{\Mkk^2}\,\Big( 1 - \frac{1}{N_c} \Big)\,
    \frac{(\widetilde\Delta_D)_{12}\otimes (\widetilde\Delta_D)_{12}}{\cos^2\theta} \,, 
    \quad\nonumber\\
   \widetilde C_1 &=& \frac{\pi\alpha_s L}{\Mkk^2}\,\Big( 1 - \frac{1}{N_c} \Big)\,
    \frac{(\widetilde\Delta_d)_{12}\otimes (\widetilde\Delta_d)_{12}}{\sin^2\theta} \,,
\end{eqnarray} 
\begin{align}
   C_4 &= -N_c\,C_5 = - \frac{4\pi\alpha_s L}{\Mkk^2}\,\bigg[
    \frac{r_1}{4}\,(\Delta_D)_{12}\,(\Delta_d)_{12} \nonumber\\
   &\hspace{4mm}\mbox{}+  
    \frac{(\widetilde\Delta_D)_{12}\otimes (\widetilde\varepsilon_d)_{12}}{\cos^2\theta} 
    + \frac{(\widetilde\varepsilon_D)_{12}\otimes(\widetilde\Delta_d)_{12}}{\sin^2\theta} 
    \bigg] \,, \nonumber
\end{align}
where $\Delta_i,\,\widetilde\Delta_i=\ord{(1)}$, but $r_1,\,\widetilde\varepsilon_i=\ord(v^2/\Mkk^2)$.

\begin{figure}
\begin{center}
\begin{tabular}{c}
\includegraphics[width=0.98\columnwidth]{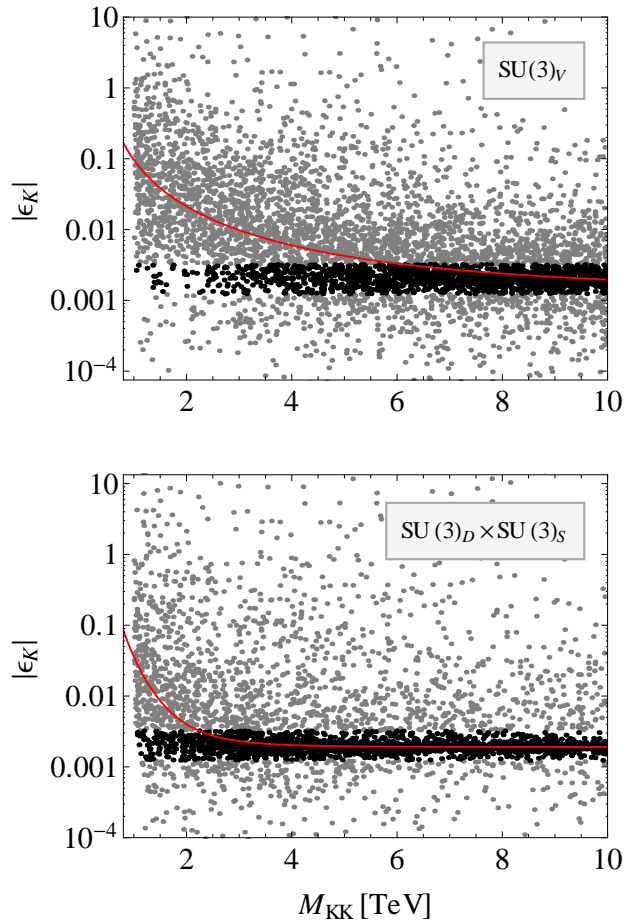}
\end{tabular}
\end{center}
\vspace{-5mm}
\caption{\label{fig:epsK} 
The CP-violating observable $|\epsilon_K|$ as a function of $\Mkk$ for the minimal RS model (upper plot) and the model with a pseudo-axial gluon with $\tan\theta=1$ and $\xi=1$ (lower plot). The red lines represent fourth-order polynomial fits in $\Mkk^{-1}$, which illustrate the decoupling behavior for large $\Mkk$.}
\end{figure}

We have implemented the exact results for the Wilson coefficients, including small electroweak contributions and higher-order terms in $v/\Mkk$, and computed $|\epsilon_K|$ for a dataset of 5000 parameter points yielding acceptable values for the quark masses and CKM parameters. The results are shown in Figure~\ref{fig:epsK}. The upper plot corresponds to the minimal RS model, while the lower one is obtained in the extended model with $\tan\theta=1$ and $\xi=1$. The results are largely independent of the precise values of these parameters as long as $\sin\theta$ or $\cos\theta$ are not very small. The black points correspond to parameter sets which reproduce $|\epsilon_K|\in [1.2,3.2]\cdot 10^{-3}$ within the phenomenologically allowed range, taking into account the uncertainty in the SM prediction. The solid lines in the two plots show fits to the median value of $|\epsilon_K|$, from which one can read off the resulting bounds on the KK scale. While in the minimal RS model the median falls within the allowed range only for large values $\Mkk>9$\,TeV, in the extended model this happens already for $\Mkk>2.5$\,TeV. Note that the masses of the lowest-lying KK states are approximately equal to $2.45\,\Mkk$. The following table shows the percentages of parameter sets which reproduce $|\epsilon_K|\in [1.2,3.2]\cdot 10^{-3}$ for different $\Mkk$ bins and three values of $\xi$:

\begin{table}[h!]
\begin{tabular}{c|c|c|c|c|c}
~$\Mkk$\,[TeV]~ & ~1--2~ & ~2--3~ & ~3--4~ & ~4--5~ & ~5--10~ \\
\hline
~original model~ & \phantom{1}5\% & \phantom{1}10\% & 18\% & 25\% & 41\% \\
~extended model ($\xi=0.5$)~ & 20\% & 43\% & 59\% & 66\% & 77\% \\
~extended model ($\xi=1$)~ & 18\% & 38\% & 56\% & 64\% & 77\% \\
~extended model ($\xi=2$)~ & \phantom{1}14\% & 32\% & 50\% & 62\% & 76\%
\end{tabular}
\vspace{-1mm}
\end{table}

\noindent
Even for low KK scales the RS flavor problem is solved in the extended model, since no significant fine-tuning is required to keep $|\epsilon_K|$ within the allowed range. 

A characteristic prediction of the minimal RS model with bulk gauge fields is the existence of a tower of KK gluons, whose lowest-lying mode has a mass given by $m_{g_1}\approx 2.448\,\Mkk$ \cite{Davoudiasl:1999tf}. In our extended model an additional tower of KK pseudo-axial gluons with mixed-chirality couplings to SM quarks appears. Their masses are very similar to the masses of the KK gluons. In addition, our model gives rise to massive, color-octet scalar fields, which in the scenario sketched above would transform as doublets under $SU(2)_L$. It also follows that quarks and leptons must couple to different scalar fields, which may offer a possibility to explain their different mass scales. A detailed discussion of the extended scalar sector, and of the collider and flavor phenomenology of our model, will be presented elsewhere.

With the RS flavor problem solved, the strongest constraints on the KK scale of the extended RS model come from the electroweak sector, namely from the oblique parameters $S$, $T$ and the $Zb\bar b$ couplings. The corresponding bounds can be relaxed by extending the electroweak gauge group in the bulk to include a custodial symmetry \cite{Agashe:2003zs,Agashe:2006at,Djouadi:2006rk}. If this is done the $T$ parameter is protected, but at tree level the constraint from the $S$ parameter requires that $\Mkk>2.4$\,TeV \cite{Casagrande:2008hr}. Lowering the scale factor $L$ of the extra dimension \cite{Davoudiasl:2008hx}, replacing the IR brane by a soft wall \cite{Batell:2008me}, or deforming the metric near the IR brane \cite{Cabrer:2011fb} provide alternatives for relaxing these bounds. On the other hand, depending on the exact realization of the extended scalar sector, new contributions to electroweak precision observables may arise in our model.

In summary, we have shown that an extension of the strong-interaction gauge group in the bulk provides a natural and in many ways minimal solution to the flavor problem of warped extra-dimension models. The most severe flavor constraint, arising from the $\epsilon_K$ parameter in $K$--$\bar K$ mixing, is relaxed by an exact cancellation of the KK gluon and pseudo-axial gluon contributions to the dangerous mixed-chirality four-quark operators at leading order in $1/\Mkk^2$, leaving only tolerable $v^2/\Mkk^4$ contributions to these operators. As a result, KK scales as low as 1--2\,TeV, corresponding to KK excitations of the gluon and pseudo-axial gluon as light as 2.5--5\,TeV, become allowed without significant fine-tuning. From the point of view of LHC collider physics, the extended RS model is thus more promising than the original one.

\vspace{1mm}
{\em Acknowledgements:\/} 
We are grateful to K.~Agashe and L.~Da Rold for pointing out the relevance of the extended scalar sector, and to U.~Haisch and S.~Westhoff for useful discussions. This work was supported by the Federal Ministry for Education and Research grant 05H09UME.

\end{document}